\documentclass[11pt, a4paper]{amsart}

\usepackage{amsmath}
\usepackage{latexsym}
\usepackage[all]{xy}
\usepackage{color}

\include{amslatex}

\numberwithin{equation}{section}
\begin{document}

\title[$\mathcal{C}$, $\mathcal{P}$, $\mathcal{T}$ and charged particle dynamics]{ { $\mathcal{C}$, $\mathcal{P}$, $\mathcal{T}$ operations and classical point charged particle dynamics}}

\maketitle
\author{

\begin{center}

Ricardo Gallego Torrom\'{e}\footnote{email: rigato39@gmail.com}\\
Departamento de Matem\'atica\\
 Universidade Federal de S\~ao Carlos, Brazil
\end{center}}
\begin{abstract}
The action of parity inversion, time inversion and charge conjugation operations on several differential equations for a classical point charged particle are described. Moreover, we consider the notion of {\it symmetrized acceleration} $\Delta_q$ that for models of point charged electrodynamics is sensitive to deviations from the standard Lorentz force equation. It is shown that $\Delta_q$ can be observed with current or near future technology and that it is an  useful quantity for probing radiation reaction models. To illustrate these points we consider four different models for the dynamics of point charged particles and radiation reaction.
\end{abstract}

\section{Introduction}

It is of relevance in electrodynamics to have a consistent theory of radiation reaction of charged particles and in particular, a consistent classical model of dynamics. This is important  not only from a theoretical perspective, but also for applications in accelerator science and laser-plasma acceleration, domains where modern applications of electrodynamics are reaching dynamical regimens where radiation reaction effects are of relevance \cite{D, Vranic}. There, to have a consistent mathematical theory for radiation reaction is necessary for further technical developments.

However, the theory of radiation reaction of charged particles is severely challenged by the fact that there are very few physical systems capable to probe experimental signatures of radiation reaction effects from individual charged particles. {\it Penning traps} \cite{Spohn1, Spohn2} and high intensity laser-plasma systems \cite{Vranic} are between these few systems where individual traces of radiation reaction are observable. Therefore, any other possible observable effect for the radiation reaction will be of relevance to clarify which models are excluded experimentally.

We are interested in radiation reaction models for classical point charged particles. The reason for this is that many of the applications where radiation reaction is of practical interest are in the classical regime, where particles have well defined classical trajectory. Think for example in charged particles composing a bunch in the beam in a storage ring or circular accelerator. The description for such systems is purely classical. In order to describe such systems, one needs a consistent classical model for radiation reaction. 

We have also assume  that the particles are point particles. This is a simple assumption, made for instance by Dirac in his theory of the classical electron \cite{Dirac} and that leads to testable dynamical models. Thus, simplicity and predictive power are reasons to adopt the idealized description of point particle. Otherwise, one needs to deal with rather involved models with {\it Poincar\'e tensions} that, because their complicated structure, will be rather difficult to analyzed as we will make below.

In this paper we show how the parity inversion $\mathcal{P}$ and the time inversion $\mathcal{T}$ are defined at the classical level and how they act on several models of point charge dynamics. We will also consider a charge conjugation operation  $\mathcal{C}$  and the composed transformation operations $\mathcal{CP}$, $\mathcal{PT}$, $\mathcal{CT}$ and $\mathcal{CPT}$. Different equations have different behavior respect to the action of such operations. This has two consequences. Firstly, it provides a direct relation with the corresponding possible underlying quantum theory, based on the principle that the symmetries of the quantum theory must be preserved by the corresponding classical limit. As a consequence, it turns out that some of the models considered below are consistent with an underlaying relativistic quantum theory, while others are not.  Secondly, we will show how the study of these operations suggests to define a quantity that we have called {\it symmetrized acceleration},  which is sensitive to the radiation reaction effects and that  is observable with current technology. This implies the possibility to falsify the predictions of several models of point charged particle and to compare some models with others.

The structure of this paper is the following.
 We first define the (classical) operations {$\mathcal{P}$, $\mathcal{T}$  $\mathcal{C}$ acting on the electromagnetic fields. For all the models considered below, electromagnetic fields will be either 1. Solutions of the Maxwell's theory for smooth external sources or, 2. Solutions of alternative theories of electrodynamics such that the electric and magnetic field bare in common the same transformation properties under the action of {$\mathcal{P}$, $\mathcal{T}$ and $\mathcal{C}$ as in Maxwell's theory. Then we consider how this operations act on other physical electrodynamic quantities, like potentials and densities. After this,  it is shown how the operations $\mathcal{P}$, $\mathcal{T}$ and $\mathcal{C}$ act on several equations of motion for the  point charged particle. In particular, we analyze the case of  the Lorentz force equation \cite{Jackson}, Lorentz-Dirac equation \cite{Jackson, Dirac}, Landau-Lifshitz equation \cite{LandauLifshitz, Spohn2}, Bonnor-Larmor's \cite{Bonnor, Larmor} equation and a recently equation proposed by the author in the framework of higher order jet electrodynamics  \cite{Ricardo012, Ricardo012b}. Although this list is far from exhausting all the models found in the literature to describe radiation reaction, they provide an illustration and certainly several insights on the relation between the description of radiation reaction and {$\mathcal{P}$, $\mathcal{T}$ and $\mathcal{C}$. Two examples are the following. First, in order to be the classical limit of a relativistic quantum field theory, a classical model of point electrodynamics must be invariant under $\mathcal{CPT}$. It turns out that while the Lorentz-Dirac, Landau-Lifshitz and the equation related with higher order electrodynamics are  invariant under  $\mathcal{CPT}$, while Bonnor-Larmor's theory is not. The second general conclusion that seems to be extracted from studying the above symmetries is that models of radiation reaction breaks $\mathcal{T}$. That is, radiation-reaction dynamics is irreversible.
 After this, the {\it symmetrized four-acceleration} $\Delta^\mu_q=\ddot{x}^\mu_q+\ddot{x}^\mu_{-q},\,\mu=0,1,2,3$ for each of the models is considered. $\Delta^\mu_q$ measures the difference in the dynamics between a charged particle and the corresponding charged anti-particle under the influence of the same external electromagnetic field. $\Delta^\mu_q$ is sensitive to the specific model of radiation reaction, being in general different from zero. We illustrate this with the theoretical evaluation of $\Delta^\mu_q$ for the above models. It will be shown that for some models $\Delta^\mu_q$ can be observed with current or with projected facilities. We expect a non-trivial effect for Landau-Lifshitz equation and for the equation introduced by the author \cite{Ricardo012, Ricardo012b}. Then we discuss several experimental possibilities where $\Delta^\mu_q$ can be used to probe radiation reaction models. While from the methodological point of view that we will describe the Lorentz-Dirac and Landau-Lifshitz models give the equivalent results, the method can be used to discriminate these two models from the Bonnor-Larmor's model and from the model introduced by the author. A brief discussion is set at the end of the paper which includes a comment on the violation of $\mathcal{T}$-symmetry when radiation reaction processes appear, the relation $\mathcal{CPT}$ and the underlying quantum theories and a discussion on the experimental possibilities to measure $\Delta^\mu_q$.

\section{$\mathcal{P}$, $\mathcal{T}$ and $\mathcal{C}$ transformations in Maxwell electromagnetic theory}
Let the spacetime be a Lorentzian manifold $(M,\eta)$, where $M$ is a $4$-dimensional manifold and $\eta$ is a Lorentzian metric. In a linear, homogeneous electromagnetic media, Maxwell equations can be casted in a coordinate invariant form as
\begin{align}
dF=0,\quad d\star F=J,
\label{covariantmaxwellequations}
\end{align}
where $F$ is the Faraday $2$-form, $\star$ is the Hodge's star operator of $\eta$ and $J$ is the charge current density. These equations are invariant under global isometries of the metric $\eta$. In the particular case when $\eta$ is the Minkowski metric, the isometry group is the full Lorentz group. which contains the parity inversion linear transformation $P$, the time inversion transformation $T$ and the combined $PT$ transformation. If $M$ is time oriented there is  a timelike vector field $W$ whose integral curves represent ideal observers. However, the existence of a globally defined timelike vector field unnecessarily constraints the topology of the spacetime $M$. Moreover, for the aims of the present study, it is more convenient if $W$ is  defined locally  in an open set $U_W$,  since we will deal with local properties of some ordinary differential equations. Therefore, all our considerations will be restricted to the domain of definition of $W$. $TM$ is the tangent bundle of $M$ and $TU_W=\,\bigsqcup_{x\in U}\,T_xU_W$. A vector field $V$ on $U$ will be a section of $TU_W$, $V\in \,\Gamma\, TU_W$. In particular, given a timelike vector field $W\,\in\,\Gamma \,TU_W$, we can consider the orthogonal complement   $\Pi_W(x)\subset \,T_xM$ with respect to $\eta$ of the $1$-dimensional sub-space $\langle W\rangle(x)$ generated by $W(x)$ in $T_xM$. The corresponding orthogonal bundle $\Pi_W:=\,\bigsqcup_{x\in U_W}\Pi_W(x)$; $\Gamma \, \Pi_W$ is the corresponding set of sections.
In natural units, the associated local electric and magnetic fields are defined by the relations
 \begin{align}
 {\bf E}_W=\,(\iota_W  F)^*,\quad {\bf B}_W=-(\iota_W \star F)^*,
 \end{align}
 where $\iota_W F$ is the inner product of the $2$-form $F$ with the vector field $W$ and the $*$-dual of $1$-forms is the image of the natural isomorphism induced by the spacetime metric $\eta^{-1}$ \cite{BennTucker}. It is direct that ${\bf E}_W,{\bf B}_W\in\, \Gamma\, \Pi_W$. 
 
There is a linear isomorphism between $\Gamma\, \Pi_W$ and $\Gamma\, T\mathbb{R}^3$. Thus, in very specific coordinate systems, ${\bf E}_W\mapsto \vec{E}_W$ and ${\bf B}_W\mapsto \vec{B}_W$, where $\vec{E}_W$ and $\vec{B}_W$ depend upon the parameter time determined by $W$, since the isomorphism depends on $W$ and then on the proper time parameter of $W$. It is in this sense that ${\bf E}_W,{\bf B}_W$ are the $4$-dimensional version of the electric and magnetic fields $\vec{E},\vec{B}$.  A different vector $W'$ determines different time parameter and different electric and vector fields  $\vec{E}_{W'}$ and $\vec{B}_{W'}$.   Since the vector field $W$ will be fixed in our considerations, the sub-index $W$ will be deleted from the notation.

Any particular $(1+3)$ decomposition of the equations \eqref{covariantmaxwellequations} corresponds to the Maxwell's equations for $\vec{E}$ and $\vec{B}$,
\begin{align}
\vec{\nabla}\cdot\vec{B}=0,\quad
\vec{\nabla}\times \vec{E}=-\frac{\partial \vec{B}}{\partial t},
\label{homogeneous}
\end{align}
\begin{align}
\vec{\nabla}\cdot\vec{E}=\rho,\quad
\vec{\nabla}\times \vec{B}=\,\vec{J}+\frac{\partial
\vec{E}}{\partial t }.
\label{non-homogeneous}
\end{align}

We assume the existence of two operations that we call parity inversion $\mathcal{P}$ and time inversion $\mathcal{T}$, such that the following properties hold,
 \begin{enumerate}
 \item In the case of the Minkowski space, as transformations of $TM$, both $\mathcal{P}$ and $\mathcal{T}$ coincide with the parity inversion $P$ and time inversion $T$,
 \item $\mathcal{P}$ and $\mathcal{T}$ leave invariant Maxwell equations \eqref{homogeneous} and \eqref{non-homogeneous} for all possible electric and magnetic fields $\vec{E}, \,\vec{B}$. That is, there is a definite transformation rules $(\vec{E},\vec{B})\mapsto (\mathcal{P}(\vec{E}),\mathcal{P}(\vec{B})),\,(\mathcal{T}(\vec{E}),\mathcal{T}(\vec{B}))$ such that the last two pairs are solutions of the Maxwell equations,
 \item For $Z\in TU$, let $Z^\mu\cdot\partial_\mu$ be the associated directional derivative. Then the properties of {\it Table 1}  holds good,
 \begin{center}
{\it Table 1}: Transformations of differential operators.
\begin{tabular}{|r|l|}

\hline  $\quad{\mathcal{P}}$ & $\quad{ \mathcal{T}}$\\

\hline $Z\in \langle W\rangle,\quad \mathcal{P}(Z)=\,Z $ & \,\,$Z\in \langle W\rangle,\quad \mathcal{T}(Z)=\,-Z$ \\
\hline $Z\in \Pi^-_W, \quad \mathcal{P}(Z)=\,-Z $ & $Z\in \tilde{\Pi}^+_W, \quad \mathcal{T}(Z)=\,Z$\\
\hline $Z\in \Pi^+_W,\quad \mathcal{P}(Z)=\,Z $ & $Z\in \tilde{\Pi}^-_W, \quad \mathcal{T}(Z)=\,-Z$\\
\hline \,$Z\in \langle W\rangle,\quad \mathcal{P}(Z^\mu\cdot\partial_\mu)=\,Z^\mu\cdot\partial_\mu $ & \,\,$Z\in \langle W\rangle,\quad\mathcal{T}(Z^\mu\cdot\partial_\mu)=\,-Z^\mu\cdot\partial_\mu$ \\
\hline $\,Z\in \Pi^-_W,\quad \mathcal{P}((Z^\mu\cdot\partial_\mu)=\,-Z^\mu\cdot\partial_\mu $ & $Z\in \tilde{\Pi}^-_W, \quad \mathcal{T}(Z^\mu\cdot\partial_\mu)=\,Z^\mu\cdot\partial_\mu$ \\
\hline
\end{tabular}
\end{center}
 \end{enumerate}
 where $\Pi_W=\Pi^+_W\oplus \Pi^-_W=\,\tilde{\Pi}^+_W\oplus \tilde{\Pi}^-_W$ but $\Pi^+_W\neq \tilde{\Pi}^+_W,\,\Pi^-_W\neq \tilde{\Pi}^-_W$.

These properties are not axioms for $\mathcal{P}$ and $\mathcal{T}$ but we will see that under natural physical constraints they determine uniquely the action on each electromagnetic field or quantity of relevance for electrodynamics.
Note also the following direct consequences for $\mathcal{P}$ and $\mathcal{T}$,
\begin{enumerate}
  \item Both $\mathcal{P}$ and $\mathcal{T}$ are idempotent, $\mathcal{P}^2=\mathcal{T}^2=Id,$

\item In the case of the Minkowski space, as linear operators, $\mathcal{P}^2=\mathcal{T}^2=Id,$ correspond to isometries of $\eta$.
\end{enumerate}
These two properties are direct generalizations of the corresponding $P$ and $T$ as isometries on Minkowski spacetime. In the general case, the three first rows of {\it Table 1} when defined globally on $TM$ imply that the pairs $(M,\mathcal{P})$ and $(M,\mathcal{T})$ are non-isometric almost product manifolds.

 In order to fix the actions of $\mathcal{P}$ and $\mathcal{T}$ in electromagnetic fields and associated quantities, let us consider the Lorentz  invariant $I_1=\vec{E}^2-\vec{B}^2$. It is well known that one can make a Lorentz transformation and either stay in the case where 1. $\vec{B}=\vec{0}$ or 2. $\vec{E}=\vec{0}$, depending on the sign of $I_1$. Let us consider the first case and impose $\vec{B}=\vec{0}$ in the Maxwell's equations. Then from the Lorentz invariance of the Maxwell's equations \eqref{homogeneous} and  \eqref{non-homogeneous} and the assumption that $\mathcal{P}$ corresponds to a Lorentz transformation, it follows that either
 \begin{enumerate}
 \item  $\mathcal{P}(\vec{E})=\,-\vec{E}$ and $\mathcal{P}(\rho)=\rho$ or,
 \item $\mathcal{P}(\vec{E})=\,\vec{E}$ and $\mathcal{P}(\rho)=-\rho$.
 \end{enumerate}
The second possibility is not compatible with local distributions of charge density, that is, if $\rho$ depends of one point in the spacetime. Consider a point charged particle located in the center of the origin in the intersection of a normal coordinate system of $\eta$ with $U$.  Then application of the Gauss's law for a small enough  ball ${\bf B}_2(r,0){\bf B}_2(r,0)$ implies
\begin{align*}
\int_{{\bf B}_2(r,0)} \,\vec{\nabla}\vec{E}\,d^3x=\,\int_{\mathbb{S}_2} \vec{E}\cdot d\vec{S}=\int_{{\bf B}_2(r,0)} \,\rho\,d^3x=Q,
\end{align*}
 being Q the charge of the particle. On the other hand, after acting with $\mathcal{P}$, we have
 \begin{align*}
 \int_{{\bf B}_2(r,0)} \,\big(\mathcal{P}(\vec{\nabla})\mathcal{P}(\vec{E})\big)\,d^3x=\,\,\int_{{\bf B}_2(r,0)}  \,-\vec{\nabla}\vec{E}\,d^3x=\,\,\int_{\mathbb{S}_2} -\vec{E}\cdot d\vec{S}=-Q.
 \end{align*}
By the action of parity inversion, the charge in the center of coordinates changes. This can be interpreted as a non-local action of $\mathcal{P}$ on $\rho$ and therefore, such effect is not physically acceptable, since we have assumed that $\rho$ is local. Thus, only the first option above is valid if $\rho$ is a local density.

 In a similar way, we can consider the case when $\vec{E}=\vec{0}$. Then equation \eqref{non-homogeneous} reduces to the  Amper\'e's law, that still must be invariant under the action of $\mathcal{P}$ on $\vec{B}$ and $\vec{J}$. There are two possible solutions compatible with the properties of $\mathcal{P}$,
\begin{enumerate}
\item $\mathcal{P}(\vec{B})=\vec{B}$ and $\mathcal{P}(\vec{J})=\,-\vec{J}$ or,

\item $\mathcal{P}(\vec{B})=-\vec{B}$ and  $\mathcal{P}(\vec{J})=\,\vec{J}$.
\end{enumerate}
Let us consider the specific case of a stationary current density of the form $\vec{J}=\,\rho\,\vec{v}$, with $\rho$ concentrated along a timelike curve whose coordinate velocity is $\vec{v}$.  As a consequence of $\mathcal{P}(\rho)=\rho$, we have that $\mathcal{P}(\vec{J})=\,-\vec{J}$ must hold. Note that in this case $\vec{J}$ is not a local property of matter, since it depends upon the speed $\vec{v}$ of the charged particles conforming the current.

We can analyze in a similar way the action of the $\mathcal{T}$ operation. For a current density of the form $\vec{J}=\,\rho\vec{v}$, we have that $\mathcal{T}(\vec{J})=-\vec{J}$. In the case that $\vec{B}=\vec{0}$, Maxwell's equations reduce to 
\begin{align*}
\frac{\partial \vec{E}}{\partial t}=\,-\vec{J},\quad \vec{\nabla}\vec{E}=\rho,
\end{align*}
from which it follows that invariance under $\mathcal{T}$ implies $\mathcal{T}(\vec{E})=\vec{E}$. From this invariance together with Gauss's law, it follows that $\mathcal{T}(\rho)=\rho$. Similarly, if $\vec{E}=\vec{0}$, then Ampere's law $\vec{\nabla}\times \vec{E}=\vec{J}$ and $\mathcal{T}(\vec{J})=-\vec{J}$ implies that for invariance under $\mathcal{T}$ it is necessary that $\mathcal{T}(\vec{B})=-\vec{B}$.

The charge inversion operation  $\mathcal{C}$ is partially defined by the relations
\begin{align}
\mathcal{C}(J)=\,-J,\quad\quad \mathcal{C}(\rho)=-\rho
\end{align}
together with the relations
\begin{align}
\mathcal{C}(Z\cdot \nabla)=\,\mathcal{C}(Z^\mu\, \partial_\mu)=\, Z^\mu\cdot \partial_\mu,\quad \mu=0,1,2,3,
\label{transformationCvector}
\end{align}
for any local vector field $Z\in \,\Gamma\, TU$. This implies that $\mathcal{C}(\vec{E})=\,-\vec{E}$ and $\mathcal{C}(\vec{B})=-\vec{B}.$

Because of the homogeneous Maxwell's equations \eqref{homogeneous}, it is possible to introduce scalar and vector potentials,
\begin{align}
\vec{E}=-\vec{\nabla} \phi-\,\frac{\partial \vec{A}}{\partial t}, \quad \vec{B}=\vec{\nabla}\times \,\vec{A}.
\end{align}
It is also direct how $\mathcal{C}$, $\mathcal{P}$ and $\mathcal{T}$ operates on the scalar and vector potentials, from the corresponding rule from the transformation of $\vec{E}$ and $\vec{B}$ given above, relation \eqref{transformationCvector} and {\it Table 2} below.

The discussion in this section is resumed in {\it Table 2} and  {\it Table 3}.
With such transformation rules, equations \eqref{homogeneous} and \eqref{non-homogeneous} are invariant for each of the transformations $\mathcal{P}$, $\mathcal{T}$ and $\mathcal{C}$ independently. Note that for the electromagnetic quantities, there is no other choice compatible with the physical principles of local charge distribution, invariance of the Maxwell's equations under parity inversion, time inversion, charge conjugation and the Lorentz invariance of the Maxwell's equations.
Furthermore,
following {\it Table 1}, one can classify vector fields as {\it proper vector fields} if $Z\in \Pi^-_W$ or {\it pseudo-vector fields} if $Z\in \,\Pi^+_W$. An example of proper vector field is the $4$-electric field ${\bf E}$ associated to the field $\vec{E}$, while an example of pseudo-vector field is the $4$-magnetic field ${\bf B}$ associated with the $3$-vector field $\vec{B}$. Therefore, we can say that $\vec{E}$ is a proper vector while $\vec{B}$ is a pseudo-vector $\vec{B}$ \footnote{This is equivalent to the classification as vector and axial vectors respect to $P$ \cite{Jackson}.}.

The famous $\mathcal{CPT}$ theorem asserts that any Lorentz invariant local quantum field theory with Hermitian Hamiltonian bounded from below must be invariant under the quantum symmetry $\mathcal{\hat{C}\hat{P}\hat{T}}$ \cite{StreaterWightman, WeinbergI}. As a consequence of this fact, the correspondence principle and Ehrenfest's theorem, the classical limit of a quantum equation of motion in a relativistic theory which is $\mathcal{\hat{C}\hat{P}\hat{T}}$-invariant is also invariant. The operators $\hat{\mathcal{P}}$, $\mathcal{\hat{T}}$ and $\mathcal{\hat{C}}$ have a  classical limit such that their action on the electromagnetic quantities coincide with the action of our operators and such that also the classical limit of the $\mathcal{\hat{C}\hat{P}\hat{P}}$ is trivial. Therefore, we can identify the classical operators $\mathcal{P}$, $\mathcal{T}$ and $\mathcal{P}$ as the classical limit of the corresponding quantum operators in a quantum theory that we do not specify. We do not introduce the details of the quantum theory, since the only data that we need to apply to models of point charged electrodynamics are the formal properties described in {\it Table 1}. It is this link between classical and quantum description that allow us to formulate a rule, according to which if a theory is not covariant under $\mathcal{\hat{C}\hat{P}\hat{T}}$ it is not the classical limit of a quantum relativistic local field theory.

\section{The action of $\mathcal{P}$, $\mathcal{T}$ and $\mathcal{C}$ on the equations of motion of a point charged particle}
Let us consider the Lorentz force equation for a point  particle written in the $1+3$ decomposition,
\begin{align}
m\,\ddot{\vec{x}}=\,q\,\gamma(\vec{E}+\,\dot{\vec{x}}\times \vec{B}),
\label{Lorentzlaw}
\end{align}
where $m$ is the rest mass of the particle, $q$  is the charge and $\gamma$ is the Lorentz factor $\gamma=\frac{1}{\sqrt{1-\,\dot{\vec{x}}^2}}$. In this equation, the parameter $\tau$ is the proper time of the metric $\eta$. The action of  $\mathcal{P}$, $\mathcal{T}$ and $\mathcal{C}$  on $\nabla$,  $\partial_t$, $\tau$, $\frac{d}{d\tau}$, $m$ and $q$ is given in {\it Table 3}.
\begin{center}
{\small {\it Table 2:} Transformation rules 1\quad \quad\quad{\it Table 3}: Transformation rules 2}
\end{center}
\begin{center}
\begin{tabular}{|r|l|l|l|}
\hline  & ${\mathcal{P}}$ & ${ \mathcal{T}}$ & $\mathcal{C}$\\

\hline $\,\,\,\vec{E}$ & $-\vec{E}$ & $\,\,\,\vec{E}$ & $-\vec{E} $\\
\hline $\,\,\,\vec{B}$ & $\,\,\,\vec{B}$ & $-\vec{B}$ & $-\vec{B}$\\
\hline $\,\,\,\,\rho$ & $\,\,\,\,\rho$ & $\,\,\,\,\rho$ & $-\rho$ \\
\hline $\,\,\,\vec{J}$ & $-\vec{J}$ & $-\vec{J}$ & $-\vec{J}$ \\
\hline $\,\,\,\phi$ & $\phi$ & $\,\,\,\phi$ & $-\phi $\\
\hline $\,\,\,\vec{A}$ & $-\vec{A}$ & $\,\,\,-\vec{A}$ & $-\vec{A} $\\
\hline
\end{tabular}\quad \quad \quad \quad \quad
\begin{tabular}{|r|l|l|l|}
\hline  & ${\mathcal{P}}$ & ${ \mathcal{T}}$ & $\mathcal{C}$\\
\hline $\,\,\,\vec{\nabla}$ & $-\vec{\nabla}$ & $\,\,\,\vec{\nabla}$ & $\,\,\,\vec{\nabla}$\\
\hline $\,\,\, \partial_t$ & $\,\,\,\partial_t$ & $-\partial_t$ & $\,\,\,\partial_t$\\
\hline $m$ & $m$ & $m$ & $m$\\
\hline $q$ & $q$ & $q$ & $-q$ \\
\hline $\tau$ & $\tau$ & $-\tau$ & $\tau$\\
\hline $\frac{d}{d\tau}$ & $\frac{d}{d\tau}$ & $-\frac{d}{d\tau}$ & $\frac{d}{d\tau}$\\
\hline
\end{tabular}
\end{center}
 A direct consequence from  {\it Table 2} and {\it Table 3} is that the Lorentz force equation \eqref{Lorentzlaw} is invariant under the action of $\mathcal{P}$, $\mathcal{T}$ and $\mathcal{C}$, since the way that {\it Table 2} was fixed is to make the Lorentz equation invariant under $\mathcal{C},\,\mathcal{P},\,\mathcal{T}$ partially by making the action of $\mathcal{P}$, $\mathcal{C}$ and $\mathcal{T}$ consistent with the Lorentz force equation in {\it Table 3}. It is clear that the four dimensional expression of the Lorentz force equation
\begin{align}
m\,\ddot{x}^{\mu}=\,q\,F^{\mu}\,_{\nu}\,\dot{x}^{\nu},\quad \mu,\nu=0,1,2,3
\label{Lorentzforceequation}
\end{align}
is also invariant.

The Lorentz force equation does not take into account the reaction due  to the emission of radiation. Many equations and models have been proposed in the literature to solve the problem of the radiation reaction (for very informative source of the different approaches the reader can consult for instance \cite{Erber, Rohrlichbook, Spohn2}). Investigating the symmetries of the models is a method to understand the qualitative differences between the models and this, could provide key insights for new observables. We will consider below several models, starting with the Lorentz-Dirac equation and its behavior under $\mathcal{C},\,\mathcal{P},\,\mathcal{T}$. The Lorentz-Dirac equation lays on fundamental principles of physics and it is the starting point for many other developments. We naturally follow investigating the behavior under $\mathcal{C},\,\mathcal{P},\,\mathcal{T}$ of the Landau-Lifshitz model. This is obtained from a reduction of order in the Lorentz-Dirac equation.  \cite{LandauLifshitz, Spohn1, Spohn2}. It has been intensively argued in the literature that the Landau-Lifshitz equation is the correct equation of motion for a point charged particle and indeed, it has been used in modeling beam dynamics in particle acceleration \cite{Rohrlichbook,Spohn2}. This is a strong motivation to investigate the symmetry properties of  such theory. The other two models that we will consider are much less popular. The Bonnor-Larmor's model \cite{Bonnor, Larmor} assumes that the energy radiated is taken from the mass of the charged particle. Therefore,  in Bonnor-Larmor's model the mass $m$ is variable on time. It is known that Bonnor-Larmor's equations do not contain pre-accelerated or run-away solutions. This makes Bonnor-Larmor's theory remarkable and justified to pay attention on it from the point of view of the present paper. The last model that we will consider is a new equation of motion obtained in the framework of a theory of higher order jet electrodynamics proposed in \cite{Ricardo012, Ricardo012b}. In contrast with Bonnor-Larmor's model, the theory is based upon a consistent geometric framework for classical fields and make use of a {\it geometry of maximal acceleration} \cite{Ricardo014c}, that makes the model consistent conceptually and mathematically (if one admits the renormalization of mass procedure).  Moreover, the equation is free of the difficulties of the Lorentz-Dirac equation.
\bigskip
\\
{\bf The Lorentz-Dirac equation.} This equation is the third order differential equation \cite{Dirac}
\begin{align}
m\,\ddot{x}^{\mu}=\,qF^{\mu}\,_{\nu}\,\dot{x}^{\nu}+\,\frac{2}{3} q^2\,\big(\dddot{x}^{\mu}-
(\ddot{x}^{\rho}\,\ddot{x}_{\rho})\dot{x}^{\mu}\big),\quad \ddot{x}^{\mu}\ddot{x}_{\mu}:=\ddot{x}^{\mu}\ddot{x}^{\sigma}\eta_{\mu\sigma}.
\label{lorentzdiracequation}
\end{align}
It is easy to check that the Lorentz-Dirac equation is invariant under $\mathcal{P}$. It is not invariant under $\mathcal{T}$. Indeed, under time inversion $\mathcal{T}$, equation \eqref{lorentzdiracequation} is transformed to
\begin{align*}
m\,\ddot{x}^{\mu}=\,qF^{\mu}\,_{\nu}\,\dot{x}^{\nu}-\,\frac{2}{3} q^2\,\big(\dddot{x}^{\mu}-
(\ddot{x}^{\rho}\,\ddot{x}_{\rho})\dot{x}^{\mu}\big),\quad \ddot{x}^{\mu}\ddot{x}_{\mu}:=\ddot{x}^{\mu}\ddot{x}^{\sigma}\eta_{\mu\sigma}.
\end{align*}
The Lorentz-Dirac equation is invariant under the $\mathcal{CPT}$ transformations.
\bigskip
\\
{\bf The Landau-Lifshitz equation}. This equation is obtained by a reduction of order of the Lorentz-Dirac equation by assuming that the acceleration can be first approximate by the expression
  \begin{align}
 \ddot{x}^\mu=\,\frac{q}{m}\,F^\mu\,_\nu\,\dot{x}^\nu
 \label{approximationtoequation}
 \end{align}
  in \eqref{lorentzdiracequation}.
 Since the Lorentz force equation is invariant under $\mathcal{C},\,\mathcal{P},\,\mathcal{T}$, the Landau-Lifshitz equation hereditates  the same properties under $\mathcal{P}$, $\mathcal{T}$ and $\mathcal{C}$ than the Lorentz-Dirac equation. In four dimensional formulation, the Landau-Lifshitz equation is \cite{Poisson}
\begin{align}
m\ddot{x}^\mu=\,q F^\mu_{\nu}\,\dot{x}^{\nu}+\,\frac{2}{3} q^2\big( \frac{q}{m}\dot{x}^\sigma(\partial_\sigma F_{\mu\nu})\dot{x}^\nu+\,\frac{q^2}{m^2}\,F^{\mu\alpha}F_{\alpha\nu}\dot{x}^\nu-\,
\frac{q^2}{m^2}\,(F^\alpha\,_{\sigma}\,\dot{x}^{\sigma})(F_{\alpha\nu}\,\dot{x}^{\nu})\dot{x}^\mu\big).
\label{LandauLifshitzequation}
\end{align}
This equation is invariant under $\mathcal{P}$, as it is shown after a short check. Under $\mathcal{T}$, the Landau-Lifshitz equation is not invariant and indeed the action of $\mathcal{T}$ in equation \eqref{LandauLifshitzequation} gives the expression
\begin{align*}
m\ddot{x}^\mu=\,q F^\mu_{\nu}\,\dot{x}^{\nu}+\,\frac{2}{3} q^2\big( \frac{q}{m}\dot{x}^\sigma(\partial_\sigma F_{\mu\nu})\dot{x}^\nu-\,\frac{q^2}{m^2}\,F^{\mu\alpha}F_{\alpha\nu}\dot{x}^\nu+\,
\frac{q^2}{m^2}\,(F^\alpha\,_{\sigma}\,\dot{x}^{\sigma})(F_{\alpha\nu}\,\dot{x}^{\nu})\dot{x}^\mu\big).
\end{align*}
A similar expression is obtained after the action of the charge conjugation operator $\mathcal{C}$.
Therefore, the Landau-Lifshitz equation is invariant under the combined action $\mathcal{CPT}$.
\bigskip
\\
{\bf Bonnor-Larmor's equations}.
Let us consider now the equations of motion in Bonnor-Larmor's model \cite{Bonnor}, based on a previous work from Larmor \cite{Larmor}. This theory is based on the following equations for a point charged particle,
\begin{align}
m\,\ddot{x}^{\mu}=\,qF^{\mu}\,_{\nu}\,\dot{x}^{\nu}
\label{BonnorLarmorequation1}
\end{align}
and
\begin{align}
\dot{m}=\,\frac{2}{3}q^2\,\ddot{x}^\nu\ddot{x}_\nu.
\label{BonnorLarmorequation2}
\end{align}
The combination of both equations provides a dynamical system such that the source of the energy radiated by the point particle is extracted from the rest mass $m$ of the particle. We consider the transformation properties of \eqref{BonnorLarmorequation1} and \eqref{BonnorLarmorequation2} separately.
Formally, equation \eqref{BonnorLarmorequation1}  is the Lorentz force equation, but with a variable mass function $m(\tau)$. Both equations \eqref{BonnorLarmorequation1} and \eqref{BonnorLarmorequation2} are invariant under parity inversion $\mathcal{P}$. However, Bonnor-Larmor's theory cannot be invariant under the action of $\mathcal{T}$. This is in contrast with the Lorentz force equation.  Indeed, by {\it Table 3}, if $m(\tau)$ is invariant under time inversion operation $\mathcal{T}$, then equation \eqref{BonnorLarmorequation2} is not consistent, since the right-hand side is invariant but the left side is not: after applying to \eqref{BonnorLarmorequation2} the  $\mathcal{T}$ operation, equation \eqref{BonnorLarmorequation2}  transforms to
\begin{align*}
-\frac{d m}{d\tau}=\,\frac{2}{3}q^2\,\ddot{x}^\nu\ddot{x}_\nu,
\end{align*}
while the first  Bonnor-Larmor's equation \eqref{BonnorLarmorequation1} remains invariant\footnote{If instead one considers the transformation rule $\mathcal{T}(m)=-m$, equation \eqref{BonnorLarmorequation2} is invariant, but then equation \eqref{BonnorLarmorequation1} is not invariant.}.
Under the action of $\mathcal{C}$, Bonnor-Larmor's theory is invariant. Therefore, the theory is not invariant by the action of $\mathcal{CT}$ or $\mathcal{CPT}$. By the discussion before, Bonnor-Larmor's theory cannot be the limit of a Lorentz invariant local relativistic theory with bounded Hamiltonian. This fact is understood directly is seen directly, since a variable mass cannot be the eigenvalues of a Casimir operator for the Poincare's algebra. Note how this fact contrasts  with the case of Lorentz-Dirac-Maxwell's theory, which is $\mathcal{CPT}$ invariant and therefore, could be the classical limit of a relativistic theory. The same remarks applies to the Lorentz-Dirac-Maxwell and the Landau-Lifshitz-Maxwell theories.
\bigskip
\\
{\bf Effective equation for point charged particles in higher order jet electrodynamics}.
The last example that we consider is a new equation of motion for point charged particles obtained in the framework of {\it higher order jet electrodynamics} \cite{Ricardo012, Ricardo012b}. The equation is
\begin{equation}
m\,\ddot{x}^{\mu} =\,q\,F^{\mu}\,_{\nu}\,\dot{x}^{\nu}-
\,\frac{2}{3}\,{q^2}\,\eta_{\nu\sigma}\,\ddot{x}^{\nu}\ddot{x}^{\sigma}\,\dot{x}^{\mu}.
\label{equationofmotion}
\end{equation}
This is an implicit second order differential equation. Therefore, existence and uniqueness of solutions is much harder to prove than for an ordinary differential equation. However, the same formal approximation that allows to derive Landau-Lifshitz equation from the Lorentz-Dirac equation can be used here, that allow us to prove existence and uniqueness, at least locally \cite{Ricardo012, Ricardo012b}.
As Bonnor-Larmor's equation, equation \eqref{equationofmotion} does not have pre-accelerated and run-away solutions, which we believe are remarkable properties of the equation.

Equation \eqref{equationofmotion} is based upon the new framework of {\it higher order jet field theory}. Given a test particle represented by a world line $\alpha:I\to M$, a higher order jet field of order $k$ measured or tested by $\alpha$ is a tensor field along $\alpha$ with values on higher order $k$-jet lift of $\alpha$. For instance, a $2$-jet tensor $T$ of type $(3,0)$ will assign to each triplet $(X_1,X_2,X_3)$  of vector fields along $\alpha$ a function along $\alpha$ of the form $T_{(\alpha,\alpha',\alpha'')}(X_1,X_2,X_3)$, where in this case the parameter of the curve $\alpha$ has not been specified. The {\it raison d'$\hat{e}$tre} of this construction is that it captures the essence of radiation reaction systems, since the field itself depends upon the motion of the test particle $\alpha$. Therefore, in higher order jet field theory the notions of test particle and external field are substituted by the most adequate notion of higher order field.

 A  relevant example of higher order jet tensor is the {\it metric of maximal acceleration}. A metric of maximal acceleration assigns to each timelike curve $\alpha:I\to M$ respect to the Lorentzian metric $\eta$ a proper time parameter \cite{Ricardo014c}
\begin{align}
\tau[t]=\int^t_{t_0} \,\Big(1- \frac{\eta(D_{\alpha'}\alpha'(s),
D_{\alpha'}\alpha'(s))}{A^2 _{max}}\Big)^{\frac{1}{2}}\,ds.
\label{propertimeg}
\end{align}
 where $D$ is the covariant derivative of the Lorentzian metric $\eta$, $\alpha'$ is the velocity vector along $\alpha$ parameterized by the proper time of $\eta$ and $A_{max}$ the value of an hypothetical maximal acceleration. It is an assumption underlaying our theory that the physical proper time is associated with a metric of maximal acceleration, whose components are given by
 \begin{align}
g_{\mu\nu}(\,^2\alpha):=\,\Big(1+ \frac{  \eta(D_{\alpha'}\alpha'(s),
D_{\alpha'}\alpha'(s))}{A^2 _{max}\,\eta(\alpha',\alpha')}\Big)\eta_{\mu\nu},\quad \mu,\nu=1,...,4,
\label{maximalaccelerationmetric}
\end{align}
where $^2\alpha:I\to TTM,\,s\mapsto\,(\alpha(s),\alpha'(s),\alpha''(s))$.  This geometric structures, have different kinematical
constraints than a Lorentzian metric, in particular $g(\alpha',\alpha'')$ is of order $\frac{\eta(D_{\alpha'}\alpha'(s),
D_{\alpha'}\alpha'(s))}{A^2 _{max}}$, which is small for system well below the maximal acceleration regime but it is not zero, in contrast with a Lorentizian metric which is zero. That $g(\alpha',\alpha'')\neq 0$ has important implications for the dynamics of point charged particles in higher order jet electrodynamics and was used in the derivation of the  equation of motion \eqref{equationofmotion} as an effective description \cite{Ricardo012, Ricardo012b}.  Indeed, equation \eqref{equationofmotion} the time derivatives are taken respect to such parameter.

 Equation \eqref{equationofmotion} is invariant under $\mathcal{P}$ but not under $\mathcal{T}$ and $\mathcal{C}$. Indeed, by the action of $\mathcal{T}$ it transforms to the equation
\begin{align*}
m\,\ddot{x}^{\mu}=\,qF^{\mu}\,_{\nu}\,\dot{x}^{\nu}+
\,\frac{2}{3}\,{q^2}(\ddot{x}^{\nu}\,\ddot{x}_{\nu})\dot{x}^{\mu}.
\end{align*}
The same relation holds after the action of the time inversion $\mathcal{C}$. However, the equation is invariant by the action of the transformation $\mathcal{CPT}$.
\bigskip
\\
\section{Symmetrized acceleration}
We  note (see {\it Table 4}) that the invariance properties of the Lorentz-Dirac, Landau-Lifshitz and  equation \eqref{equationofmotion} are the same. This makes rather difficult to distinguish between these models by studying the experimental qualitative effect of the symmetries $\mathcal{C}$, $\mathcal{P}$ and $\mathcal{T}$ alone. We introduce here a quantity that allows to investigate further symmetric properties of relativistic dynamical systems.
If $\ddot{x}_q$ is the acceleration appearing in the Lorentz force law under the fixed external electric and magnetic fields $\vec{E}$ and $\vec{B}$ of a particle of mass and charge $(m,q)$ and  $\ddot{x}_{-q}$ is the acceleration for a particle $(m,-q)$ under the same external fields $\vec{E}$ and $\vec{B}$, one has the relation
\begin{align}
\Delta^\mu_q(x):=\ddot{x}^\mu_q+\ddot{x}^\mu_{-q}=0.
\label{symmetricrelationq-q}
\end{align}
Note that this function does not involves charged conjugation operation $\mathcal{C}$, since under charge conjugation \eqref{Lorentzforceequation} is strictly invariant. In a general radiation reaction model of classical point particles we call this function $\Delta^\mu_q(x)$  {\it symmetrized acceleration}.

 The physical meaning of the relation \eqref{symmetricrelationq-q} is that for a given external electromagnetic field described by the Faraday $2$-form $F$, if the world line of a particle of mass and charge $(m,q)$ is a solution of the Lorentz force equation \eqref{Lorentzforceequation} with acceleration $\ddot{x}_q$, then the particle with $(m,-q)$ will follow a world line with exactly opposite acceleration $\ddot{x}_{-q}=\,-\ddot{x}_q$ and where the initial conditions are also changed according to $\mathcal{P}$ and $\mathcal{T}$ on the initial conditions.

 We evaluate below the symmetrized acceleration for the four models of radiation reaction discussed in the previous section.
\bigskip
\\
{\bf Lorentz-Dirac.}
The corresponding symmetrized four-acceleration analogue of equation \eqref{symmetricrelationq-q} for the Lorentz-Dirac equation is
\begin{align}
\Delta^\mu_q(x)=\ddot{x}^\mu_q+\ddot{x}^\mu_{-q}=\,\frac{4}{3} \, \frac{q^2}{m}\,\big(\dddot{x}^{\mu}-
(\ddot{x}^{\rho}\,\ddot{x}_{\rho})\dot{x}^{\mu}\big).
\label{symmetricombinationLD}
\end{align}
\bigskip
\\
{\bf Landau-Lifshitz.}
The symmetrized $4$-acceleration for the Landau-Lifshitz equation is
\begin{align}
\Delta^\mu_q(x)=\,\ddot{x}^\mu_q+\ddot{x}^\mu_{-q}=\,\frac{4}{3} \frac{q^2}{m}\,\frac{q^2}{m^2}\big(\,F^{\mu\alpha}F_{\alpha\nu}\dot{x}^\nu-\,
\,(F^\alpha\,_{\sigma}\,\dot{x}^{\sigma})(F_{\alpha\nu}\,\dot{x}^{\nu})\dot{x}^\mu\big).
\label{symmetricombinationLL}
\end{align}
\bigskip
\\
{\bf Bonnor-Larmor.}
For the Bonnor-Larmor's model the symmetrized acceleration is
\begin{align}
\Delta^\mu_q(x)=\,\ddot{x}^\mu_q+\ddot{x}^\mu_{-q}=\,0.
\label{symmetriccombinationforBonnor}
\end{align}
This is a qualitatively different behaviour than for the Lorentz-Dirac and the Landau-Lifshitz theories, given by the relations \eqref{symmetricombinationLD} and \eqref{symmetricombinationLL} respectively, since in such theories the symmetrized acceleration is strictly different from zero.
\bigskip
\\
{\bf Effective equation \eqref{equationofmotion}.}
The symmetrized four-acceleration is in this case given by the expression
\begin{align}
\Delta^\mu_q(x)=\ddot{x}^\mu_q+\ddot{x}^\mu_{-q}=\,-\frac{4}{3} \,\frac{q^2}{m}\,
(\ddot{x}^{\nu}\,\ddot{x}_{\nu})\dot{x}^{\mu}.
\label{symmetricombinationnew0}
\end{align}
 It is convenient to consider the approximation \eqref{approximationtoequation}.
Then one obtains a second order ordinary differential equation
\begin{align}
\ddot{x}^\mu=\,q\dot{x}^\nu F^\sigma\,_\nu\,\big( \frac{1}{m}\delta^\mu\,_\sigma-\,\frac{2\,q^2}{3m} \,\dot{x}^\mu F_{\sigma\lambda}\dot{x}^\lambda\big).
\label{approximationofequationofmotion}
\end{align}
Theory of ODE's allows to establish local existence and uniqueness of the equation. In the limit where the approximation \eqref{approximationtoequation} remains valid, the same consequences from ODE's can be extended to the original equation.
With this approximation, the symmetrized acceleration \eqref{symmetricombinationnew0} is
\begin{align}
\Delta^\mu_q(x)=\,-\frac{4}{3} \frac{q^2}{m}\,\frac{q^2}{m^2}
(F^\mu\,_\sigma\,\dot{x}^\sigma)\,(F_{\mu\lambda}\,\dot{x}^\lambda)\dot{x}^{\mu}.
\label{symmetricombinationnew}
\end{align}
\begin{center}
{{\it Table 4:} $\Delta^\mu_q$ and discrete symmetries for  models of point charged particle}
\paragraph{}
\begin{tabular}{|r|r|r|l|l|l|l|l|}
\hline  &  Lorentz & Lorentz-Dirac & Landau-Lifshitz & Bonnor-Larmor & Higher order fields  \\
\hline $\Delta^\mu_q$  & 0 & $\frac{4}{3} \, \frac{q^2}{m}\,\big(S^\mu_{LD}-S^\mu\big)$ & $\frac{4}{3}\, \frac{q^2}{m}\,\frac{q^2}{m^2}\big(T^\mu_{LL}\big)$ & 0 & $-\frac{4}{3} \, \frac{q^2}{m}\,S^\mu $ \\
\hline $\mathcal{C}$ & YES & NO & NO & YES & NO \\
\hline $\mathcal{P}$ & YES & YES & YES & YES & YES \\
\hline $\mathcal{T}$ & YES & NO & NO & NO & NO \\
\hline $\mathcal{CP}$ & YES & YES & YES & NO & YES \\
\hline $\mathcal{PT}$ & YES & YES & YES & NO & YES \\
\hline $\mathcal{CPT}$ & YES & YES & YES & NO & YES \\
\hline
\end{tabular}
\end{center}
with
\begin{align*}
 T^\mu_{LL}:=\big(\,F^{\mu\alpha}F_{\alpha\nu}\dot{x}^\nu-\,
\,(F^\alpha\,_{\sigma}\,\dot{x}^{\sigma})(F_{\alpha\nu}\,\dot{x}^{\nu})\dot{x}^\mu\big),\quad
 S^\mu_{LD}= \,\dddot{x}^{\mu},\quad
S^\mu=\,(\ddot{x}^{\rho}\,\ddot{x}_{\rho})\dot{x}^{\mu}.
\end{align*}

We would like to make two remarks. 
 The first is that for all the models considered describing radiation-reaction imply violation of $\mathcal{T}$-invariance, in contrast with the Lorentz force equation \eqref{Lorentzforceequation} which is invariant under $\mathcal{T}$ but does not take into account radiation reaction. Such universality suggests that radiation reaction is an non-reversible process.
The second remark is that for the  Bonnor-Larmor's equation $\Delta^\mu_q$ is identically zero. Therefore, an experimental measurement of $\Delta^\mu_q$ should falsify Bonnor-Larmor's model and an absence of a non-zero $\Delta^\mu_q$ should falsify the Lorentz-Dirac, Landau-Lifshitz theory and equation \eqref{equationofmotion}. In the next {\it section} we discuss several scenarios where $\Delta^\mu_q$ could be in principle measured, which provides the possibility to contrast against experiment several models of radiation reaction of point charged particles.

\section{Examples of symmetrized four-acceleration.}
 In this section we discuss how symmetrized acceleration $\Delta^\mu_q(x)$ can be used to discriminate between the models of dynamics of point charged particle discussed above. Let us consider the $1+3$ decomposition and apply it to the relation given by the expression \eqref{symmetricombinationnew} for the model based on the equation \eqref{equationofmotion}. In this case and after using the approximation \eqref{approximationtoequation}, we obtain for the symmetrized acceleration
\begin{align}
\vec{\Delta}_q(\vec{x})=\,2\,\frac{\epsilon}{c^2}\,\frac{q^2}{m^2}\,\gamma^2 (\vec{E}+\,\dot{\vec{x}}\times \vec{B})^2\,{\dot{\vec{x}}},
\end{align}
where $c$ is the speed of light and $\epsilon=\frac{2}{3}\frac{q^2}{mc^3}$ is the {\it characteristic time} associated with the classical radius of a point charged particle \cite{Rohrlichbook, Spohn2}. For an electron or positron it has the value $\epsilon=\,0.62\times 10^{-23}s$. However, for particles or systems with $m=\,N m_e$ and $q=\,N q_e$,  the corresponding $\epsilon$ is  N times larger than for an electron.
We will consider first electrons and positrons in the following two cases.
\bigskip
\\
 {\bf There is only  magnetic field.} In this case, we have an expression of the form
\begin{align*}
|\vec{\Delta}_q(\vec{x})|(\vec{0},\vec{B})=\,2\,\frac{\epsilon}{c^2}\,\frac{q^2}{m^2}\,\gamma^2 \,(\dot{\vec{x}}\times \vec{B})^2\,|{\dot{\vec{x}}}|.
\end{align*}
The ratio with the absolute value of the Lorentz force acceleration $3$-vector is given by the expression
\begin{align}
\mathcal{R}_q (\vec{0},\vec{B}):=\,\frac{|\vec{\Delta}_q(\vec{x})|}{|\ddot{\vec{x}}|}=\,2\,\frac{\epsilon}{c^2}\,\frac{q}{m}\,\gamma\,|\dot{\vec{x}}\times \vec{B}||{\dot{\vec{x}}}|+\,\mathcal{O}(\epsilon^2),
\label{quotientfornewB}
\end{align}
where we have used the fact that the radiation reaction term is smaller compared with the Lorentz force term.
We can estimate the maximal values of the adimensional quantity \eqref{quotientfornewB} achievable with current technology. For this, a further simplification happens for $c=|\dot{x}|$, that corresponds to the ultra-relativistic regime. For a charged particle $q=e$ and if $|\dot{\vec{x}}\times \vec{B}|=\,|{\dot{\vec{x}}}||\vec{B}|\,\sin \theta,$ one has at leading order in $\epsilon$
\begin{align}
\mathcal{R}_e (\vec{0},\vec{B})= \,2.17\times10^{-12}\gamma\,\sin \theta \,|\vec{B}|(Tesla),
\end{align}
where in this adimensional expression the magnetic field is expressed in Tesla.
For current light synchrotron as Diamond, ESRF or ALS, the factor $\gamma$ can be of order $10^3$ \cite{Attwood}, while the magnetic fields achievable in the magnets is of order $1$ to $10$ Tesla \cite{Russenschuck}. This provides a range for the factor with maximal values of the order $\mathcal{R}_e (\vec{0},\vec{B})\simeq 10^{-9}-10^{-8}$. Values of this range for $\mathcal{R}_e (\vec{0},\vec{B})$ can be reach for Diamond beam ($\gamma=5733, \,\,|\vec{B}|\sim 1 \,T$), ALS ($\gamma=3720,\,\,|\vec{B}|\sim 1.27 \,T$) or ESRF ($\gamma=11800,\,\,|\vec{B}|\sim 0.83 \,T$). However, if super-conducting magnets are used, like in BLAZE \cite{Diamond}, then the magnetic field bending magnet can reach $14$ T, and  $\mathcal{R}_e (\vec{0},\vec{B})\sim \,10^{-7}$. This is a small difference in the behavior of electrons and positrons interacting with an strong external magnetic field $\vec{B}$, but hopefully it is detectable in the strength of $\vec{B}$ or $\gamma$ can be increased\footnote{A main difficulty is, however, how to work with positron beams in such facilities.}.
\bigskip
\\
{\bf There is only electric field.} In this case, the leading order in $\epsilon$ is given by the expression
\begin{align*}
\vec{\Delta}_q(\vec{x})(\vec{E},\vec{0})=\,2\,\frac{\epsilon}{c^2}\,\frac{q^2}{m^2}\,\gamma^2 \,\vec{E}^2\,{\dot{\vec{x}}}.
\end{align*}
Taking the ratio with the module of the  vector acceleration, one obtains
\begin{align}
\mathcal{R}_q (\vec{E},\vec{0})=\,\,2\,\frac{\epsilon}{c^2}\,\frac{q}{m}\,\gamma \,|\vec{E}|\,|{\dot{\vec{x}}}|.
\label{quotientfornewE}
\end{align}
For an electron, if we write $c=|{\dot{\vec{x}}}|$ we find that equation \eqref{quotientfornewE} is
\begin{align}
\mathcal{R}_q (\vec{E},\vec{0})=\,0.73\times 10^{-20}\,\gamma \,|\vec{E}|(Newton\cdot Coulomb^{-1}),
\end{align}
where in this adimensional expression, the electric field is expressed in $Newton\cdot Coulomb^{-1}$.

Current technology in laser-plasma acceleration provides electric fields of order $1\, GeV/cm$ and energies of $1\,Gev$ or higher for electron in plasma acceleration \cite{Katsouleas, Leemans}. For such fields and kinematical conditions, one has that $\mathcal{R}_q (\vec{E},\vec{0})\simeq  0.73\times 10^{-7}$. In principle, this difference is large enough to be observed with current  or future laser-plasma acceleration technology. Compare with this figures, electron accelerators could achieve gamma factors of order $10^3$, but the electric fields that electrons are exposed is much less than in laser-plasma accelerators, of order $50 MeV/m$. Therefore, the figures for $\mathcal{R}_q (\vec{E},\vec{0})$ in electron accelerators are of order $10^{-1}$ to $ 10^{-2}$ smaller than in laser-plasma acceleration systems.
\bigskip
\\
{\bf Symmetry acceleration in the non-relativistic limit}.
Note that $T_{LL}$ is of the same order than $S_{LD}-S$, since the Landau-Lifshitz equation is obtained by a reduction of order from the Lorentz-Dirac equation neglecting higher order terms in $\epsilon$. In the case of static fields, $\Delta^\mu_q$ coincides for the Lorentz-Dirac and for the equation \eqref{equationofmotion}. Moreover, it can be seen that in the ultra-relativistic limit $\gamma>>1$ $T_{LL}$ and $S$ coincide in the leading term on $\gamma$ in the case when $\vec{B}=0$. In such case $\Delta^\mu_q$ cannot discriminate between the Lorentz-Dirac, Landau-Lifshitz and the  equation \eqref{equationofmotion} at a qualitative level by the method discussed.

However, $T_{LL}$ and $S_{LD}$ can be zero in the non-relativistic regime and for particular combinations of electric and magnetic fields. This contrast with $S$, which is always different than zero for an accelerated charged particle.

An example where it could be certainly possible to see differences between $T_{LL}$ and $S$ is the following. In the slow motion regime where $\gamma\simeq 1$ and consider an  electric field $\vec{E}$ and that the charged particle has speed $\vec{v}$ parallel to $\vec{E}$. Then consider a charged particle with parameters $(q,m)=\,N(e,m_e)$ with $N>>1$ integer and $(e,m_e)$ the mass and charge of the electron. Then we have the relation
\begin{align}
\mathcal{R}_{qLL}=0,\quad \mathcal{R}_{q}=\,N\,\frac{|\dot{x}|}{c}\,0.73\times 10^{-20}\,\,|\vec{E}|(Newton\cdot Coulomb^{-1})
\label{evaluationlowenergy}
\end{align}
because $\epsilon=\,\frac{2}{3}\frac{q^2}{mc^3}$ and
where $\mathcal{R}_{qLL}$ is the symmetrized acceleration function associated to the Landau-Lifshitz equation. This possibility make use that $N$ can be large.  On the other hand, note that the quotient $\frac{|\dot{x}|}{c}$ is small, while the field is limited to values of order $10^8\,Newton\cdot Coulomb^{-1}$. Otherwise the electron can accelerated substantially and  $\gamma$ can be comparable to $1$. Thus, there are different factors that influence the observability of $\mathcal{R}$ in different directions, although the hope is that the effect is large enough to be observable, in comparison with the Landau-Lifshitz identically null effect.

\section{Discussion}

By directly investigating $\mathcal{P}$, $\mathcal{T}$ and $\mathcal{C}$ at the classical level for point charged electrodynamics, we have seen the similarity on the behavior of the models of radiation reaction. In particular, although the specific form in which $\mathcal{T}$ is broken depends on the details of the model, we recognized an universal violation of $\mathcal{T}$-invariance associated with the processes of absorbtion and emission of radiation requires a energy momentum balance treatment of the energy lost by radiation at the infinite. Note that this is a purely classical effect.

Special attention has been paid to the $\mathcal{CPT}$ symmetry. Our study indicates that some of the most relevant models in the literature of classical point electrodynamics are $\mathcal{CPT}$ invariant. This should not be a surprise, since quantum electrodynamics is $\mathcal{\hat{C}\hat{P}\hat{T}}$-invariant and therefore, we should find that possible classical limits are also $\mathcal{CPT}$-invariant.

On the other hand, the similarity in {\it Table 4} for several models of radiation reaction reveals that only the knowledge of the way $\mathcal{P}$, $\mathcal{T}$ and $\mathcal{C}$ act it is not possible to discriminate between them.
We have shown that the analysis of the symmetrized acceleration $\Delta^\mu_e$ provides a theoretical tool to investigate the qualitative behavior of models for radiation reaction and that in special regimes, the symmetrized acceleration can be currently experimentally observed. The symmetrized acceleration depends on the model in such a way that some of them can be falsified  with current laser-plasma technology. Thus,  qualitative differences for the Lorentz-Dirac, Landau-Lifshitz and the equation \eqref{equationofmotion} from one side and the Lorentz and Bonnor-Larmor's model on the other is shown to lead to experimental falsifiable predictions. We showed how the numerical value of $\Delta^\mu_q$ can be magnified in the ultra-relativistic regime.

Furthermore, we have discussed  an alternative domain where $\Delta^\mu_q$ could be detectable. The regime is based on the non-relativistic limit and in magnifying the characteristic time $\epsilon$ by a factor $N$ associated with a collective behavior of a bunch of particles. Then in the non-relativistic regime $\Delta^\mu_q$ could be used to discriminate between Lorentz-Dirac and Landau-Lifshitz models from one side and equation \eqref{equationofmotion} from the other side. Although apparently this regime provides a smaller detectable signal for $\Delta^\mu_q$ than in the ultra-relativistic regime, there is the possibility to check the difference if $N$ is large enough, of the same order than the plasma bunches in particle accelerators $N\sim 10^{10}$. However, space charge effects will be dominant in the non-relativistic regime, if there is not any other force to confine the system stable under the coherent behavior under $\mathcal{E}$. 
\bigskip
\\
{\bf Acknowledgements} R. G. T.  was supported by the Riemann Center for Geometry and Physics and by PNPD-CAPES n. 2265/2011, Brazil.

\small{
}

\end{document}